\documentclass[epjCONF]{svjour}
\usepackage{graphicx}
\usepackage[varg]{txfonts} 
\usepackage[latin1]{inputenc}
\session-title{%
Three-body break-up in deuteron-deuteron scattering at 65~MeV/nucleon
}
\begin{document}
\title{%
Three-body break-up in deuteron-deuteron scattering at 65~MeV/nucleon 
}%
\author{%
A.~Ramazani-Moghaddam-Arani\inst{1,2}\fnmsep\thanks{\email{ramazani@kvi.nl}} 
\and %
H.R.~Amir-Ahmadi\inst{1} 
\and %
A.D.~Bacher\inst{3} 
\and %
C.D.~Bailey\inst{3}
\and %
A.~Biegun\inst{1}
\and %
M.~Eslami-Kalantari\inst{1,4}
\and %
I.~Ga\v{s}pari\'c\inst{5}
\and %
L.~Joulaeizadeh\inst{1}
\and %
N.~Kalantar-Nayestanaki\inst{1}
\and %
St.~Kistryn\inst{6}
\and %
A.~Kozela\inst{7}
\and %
H.~Mardanpour\inst{1}
\and %
J.G.~Messchendorp\inst{1}
\and %
A.M.~Micherdzinska\inst{8}
\and %
H.~Moeini\inst{1}
\and %
 S.V.~Shende\inst{1}
\and %
 E.~Stephan\inst{9}
\and %
E.J.~Stephenson\inst{3}
\and %
R.~Sworst\inst{6}
}
\institute{%
KVI, University of Groningen, Groningen, The Netherlands
\and %
 Department of Physics, Faculty of Science, University of Kashan, Kashan, Iran
\and %
Indiana University Cyclotron Facility, Bloomington, Indiana, USA
\and %
Department of Physics, Faculty of Science, Yazd University, Yazd, Iran
\and %
Rudjer Bo\v{s}kovi\'c Institute, Zagreb, Croatia
\and %
Institute of Physics, Jagiellonian University, Cracow, Poland
\and %
 Henryk Niewodnicza\'nski Institute of Nuclear Physics, Cracow, Poland
\and %
University of Winnipeg, Winnipeg, Canada
\and %
Institute of Physics, University of Silesia, Katowice, Poland
}
\abstract{
We successfully identified a few final states in deuteron-deuteron
scattering at 65~MeV/nucleon at KVI using a unique and advanced
detection system called BINA. This facility enabled us to perform
cross sections and polarization measurements with an excellent
statistical and systematical precision. The analysis procedure and
part of the results of the three-body break-up channel in
deuteron-deuteron scattering at 65~MeV/nucleon are presented in this
paper.  } 
\maketitle
%
%
%

\begin{sloppypar}
The physics phenomena of nuclei are for a large part understood by
considering the interaction between its building elements, the
nucleons. In 1935 Yukawa described the nucleon-nucleon (NN) force with
an exchange of massive mesons~\cite{Yukawa} similar to the
electromagnetic interaction which can be represented by an exchange of
of a massless photon.  Few phenomenological nucleon-nucleon potentials have
been derived based on Yukawa's theory and are able to reproduce more
than 4000 data points in neutron-proton and proton-proton scattering with
extremely high precision. These so-called high-quality NN potentials
are used in Faddeev~\cite{RamazaniA_Ref2,RamazaniA_Ref3} equations to
have exact solution for a three-nucleon system. Already for the
simplest three-nucleon system, the triton, an exact solution of the
three-nucleon Faddeev equations employing two-nucleon forces (2NFs)
underestimates the experimental binding
energy~\cite{RamazaniA_Ref4}, showing that 2NFs are not sufficient to
describe the three-nucleon system accurately by the presence of a
third nucleon. The existence of an additional force, the three-nucleon
(3N) interaction, was predicted by Primakov~\cite{RamazaniA_Ref5} and
was proven rigorously by a comparison between precision data and
state-of-the-art calculations. In general, adding 3NF effects to the NN
potentials gives a better agreement between cross section data in proton-deuteron scattering
and corresponding calculations~{\cite{RamazaniA_Ref6,RamazaniA_Ref7,RamazaniA_Ref8,RamazaniA_Ref9,RamazaniA_Ref10,RamazaniA_Ref11,RamazaniA_Ref12,RamazaniA_Ref13,RamazaniA_Ref14,RamazaniA_Ref15,RamazaniA_Ref16,RamazaniA_Ref17}}
, whereas a comparison for the corresponding spin observables yields various
discrepancies~{\cite{RamazaniA_Ref7,RamazaniA_Ref8,RamazaniA_Ref9,RamazaniA_Ref18,RamazaniA_Ref19,RamazaniA_Ref20,RamazaniA_Ref21,RamazaniA_Ref22}}. 
This demonstrated that spin-dependent parts of the 3NFs are poorly
understood and that more studies in this field are needed.
\end{sloppypar}

\begin{sloppypar}
The 3NF effects are in general small in the three-nucleon system. A
complementary approach is to look into systems for which the 3NF
effects are significantly enhanced in magnitude. For this, it was
proposed to study the four-nucleon system. The experimental database
in the four-nucleon system is presently poor in comparison with the
three-nucleon system.  Most of the available data were taken at very
low energies, in particular below the three-body break-up threshold of
2.2~MeV. Also, theoretical developments are evolving rapidly at low
energies~\cite{RamazaniA_Ref23,RamazaniA_Ref24,RamazaniA_Ref25,RamazaniA_Ref26},
but lag behind at higher energies. The experimental database at
intermediate energies is very
limited~{\cite{RamazaniA_Ref27,RamazaniA_Ref28,RamazaniA_Ref29}}.
This situation calls for extensive four-nucleon studies at
intermediate energies. The goal of our work was to perform a
comprehensive measurement of cross sections and spin observables in
various $d+d$ scattering processes at 65~MeV/nucleon, namely the
elastic and three-body break-up channels. With these data, we have
drastically enriched the four-nucleon scattering database. The
extensive database of spin and cross section observables in various
deuteron-deuteron scattering processes together with
precise and ab-initio calculations can potentially reveal many details
of 3NF effects.
\end{sloppypar}

Deuteron-deuteron scattering below the pion-production threshold leads to 5 
possible final states with a pure hadronic signature, namely\\ 
\noindent
\begin{enumerate}
\item Elastic channel:   $ \vec d + d \longrightarrow d + d$ ;
\item Neutron-transfer channel:  $ \vec d + d \longrightarrow p + t$ ;
\item Proton-transfer channel:   $ \vec d + d \longrightarrow n +  ^{3}\!\rm He$ ;
\item Three-body final-state break-up: $ \vec d + d \longrightarrow p + n + d$ ;
\item Four-body final-state  break-up: $ \vec d + d \longrightarrow p + n + p + n$. 
\end{enumerate}
In this work, the three-body final-state break-up in deuteron-deuteron scattering is
further referred as the three-body break-up.  The study
and identification of these final states requires an experimental
setup with specific characteristic, namely, a large phase space
coverage, a good energy and angular resolution, and the ability of particle
identification (PID). The experiment presented in this paper was
carried out using the Big Instrument for Nuclear-polarization
Analysis, BINA, which has these
requirements~\cite{RamazaniA_Ref30}. A polarized beam of deuterons
with a kinetic energy of 65~MeV/nucleon was impinged on a liquid
deuterium target. The elastic, neutron-transfer and three- and
four-body break-up channels were identified using the energy,
scattering angles and time-of-flight (TOF) information. We, only,
discuss the three-body break-up channel in this paper.

For the analysis of the three-body break-up data we measured the
energies, polar and azimuthal angles of the two coincident, emitted
particles. Using the measured variables and considering momentum and
energy conservation, all the other kinematical variables of the
reaction can be obtained unambiguously. The kinematics of the
three-body break-up reaction are determined by using the scattering
angles of the proton and the deuteron ($\theta_{d}, \theta_{p},
\phi_{12}=|\phi_{d}-\phi_{p}|$) and the correlation between their
energies presented by the kinematical curve which is called the
$S$-curve. The energies of the proton, $E_p$, and deuteron, $E_d$, were
described as a function of two new variables, $D$ and $S$. The variable
$S$ is the arc-length along the $S$-curve with the starting point chosen
arbitrarily at the point where E$_d$ is minimum and $D$ is the
distance of the ($E_p$, $E_d$) point from the kinematical curve.
\begin{figure}[!h]
\centering
\includegraphics[width=1\columnwidth,angle=0]{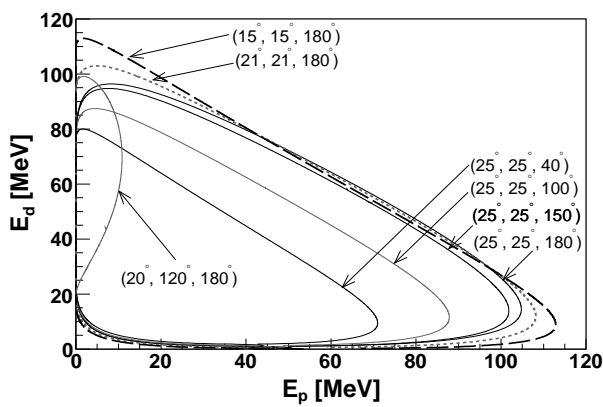}
\caption{The energy correlation between protons and deuterons
 in coincidence for the three-body break-up reaction in $\vec {d} + d$
 scattering is shown as $S$-curves for several kinematical
 configurations.  The kinematics are defined by ($\theta_{d},
 \theta_{p}, \phi_{12}$), the polar scattering angles of the proton and
 deuteron, respectively, and the relative azimuthal angle.}
\label{kinema}       
\end{figure}
The $S$-curves for several kinematical configurations are shown in
Fig.~\ref{kinema}. Each $S$-curve is labeled by three numbers. For
example, the label ($20^\circ, 30^\circ, 180^\circ$) shows a
coincidence between a deuteron that scatters to $20^{\circ}$  and a proton
that scatters to $30^{\circ}$, and the azimuthal opening angle,
$\phi_{12}$, between the deuteron and the proton is $180^{\circ}$.

\begin{sloppypar}
The first step in the event selection for the three-body break-up
channel is to find the energy correlation between the final-state 
protons and deuterons for a particular kinematical configuration ($\theta_{p},
\theta_{d}, \phi_{12}$), where $\theta_{p}$ and $\theta_{d}$
are the polar angles of the proton and the deuteron, respectively, and
$\phi_{12}$ is the difference between their azimuthal angles.  The
number of break-up events in an interval $S-\frac{\Delta S}{2}$, and
$S+\frac{\Delta S}{2}$ was obtained by projecting the events on a line
perpendicular to the $S$-curve ($D$-axis).
\begin{figure}[!h]
\centering
\includegraphics[width=1\columnwidth,angle=0]{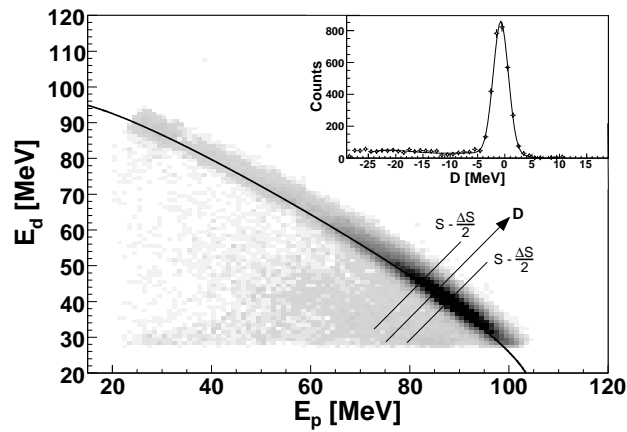}
\caption{The correlation between the energies of the deuteron and the
 proton originating from three-body break-up channel. The projection
 on the $D$-axis is shown in the inset.}
\label{25-25-180}       
\end{figure}
Figure~\ref{25-25-180} depicts the correlation between the energy of
protons and deuterons in coincidence for the kinematical
configuration, ($\theta_{p}, \theta_{d}, \phi_{12})=(25^{\circ},
25^{\circ},180^{\circ})$.  The solid curve is the expected correlation
for this configuration. One of the many $S$-intervals and the
corresponding $D$-axis are also shown. The result of the projection of
events on the $D$-axis for a particular $S$-bin is presented in the
inset of Fig.~\ref{25-25-180}. This spectrum consists of mainly
break-up events with a negligible amount of accidental
background. Particles for most of the break-up events deposit all
their energy in the scintillator, which gives rise to a peak around
zero for the variable $D$. Particles in a part of the break-up events undergo a
hadronic interaction inside the scintillator or in the material
between the target and the detector. For these events the value of $S$
is ill-defined and, therefore, considered as background (primarily to
the left-hand side of the main peak in Fig.~\ref{25-25-180}). All the
background events were subtracted by fitting a polynomial representing
the background and a Gaussian representing the signal to the projected
spectrum.  The fraction of break-up events which did not deposit their
complete energy has been estimated by a GEANT3 simulation and
corrected for when determining the cross section.
\end{sloppypar}

The interaction of a polarized beam with an unpolarized target
produces an azimuthal asymmetry in the scattering cross section. The
magnitude of this asymmetry is proportional to the product of the
polarization of the beam and an observable that is called the
analyzing power. For every kinematical point, $\xi$, the azimuthal
distribution of the scattered particles for a polarized beam is
normalized to that of the unpolarized beam. The general expression for
the cross section of any reaction induced by a polarized spin-1
projectile is~{\cite{RamazaniA_Ref31,RamazaniA_Ref32}}:
\begin{sloppypar}
\begin{eqnarray}
\sigma(\xi) = \sigma_0(\xi)\lbrack1&+&\sqrt{3}p_Z\rm {Re}(iT_{11}(\xi))\cos\phi \nonumber\\
&-&\frac{1}{\sqrt{8}}p_{ZZ}\rm {Re}(T_{20}(\xi))\nonumber\\
&-&\frac{\sqrt{3}}{2}p_{ZZ}\rm {Re}(T_{22}(\xi))\cos 2\phi \rbrack.
\label{ddbreakcrosFurmola}
\end{eqnarray}
where $\sigma$, $\sigma_{0}$ are the polarized and unpolarized cross
sections, respectively, $\xi$ represents the configuration
$(S,\theta_{p}, \theta_{d}, \phi_{12})$. Note that
Eq.~\ref{ddbreakcrosFurmola} does not contain terms with ${\rm
Im}(iT_{11})$, ${\rm Re}(T_{21})$, and ${\rm Im}(T_{21})$. These
contributions vanish because we took explicitly $\beta=90^\circ$ and
$\phi_{12}=|\phi_1 - \phi_2|$. The angle $\beta$ is the angle between
the polarization axis and the momentum of the incoming beam.  In this
work, the variables $\rm {Re}(iT_{11})$, $\rm {Re}(T_{20})$ and $\rm
{Re}(T_{22})$ will be referred to as $iT_{11}$, $T_{20}$ and $T_{22}$,
respectively.  The parameters $iT_{11}$ and $p_{Z}$ are the
vector-analyzing power and the vector beam polarization,
respectively. The observables $T_{20}$ and $T_{22}$ are the
tensor-analyzing powers, $p_{ZZ}$ is the tensor polarization of the
beam, and $\phi$ is the azimuthal scattering angle of the deuteron.
\end{sloppypar}
\begin{figure}[!h]
\centering
\includegraphics[width=1\columnwidth,angle=0]{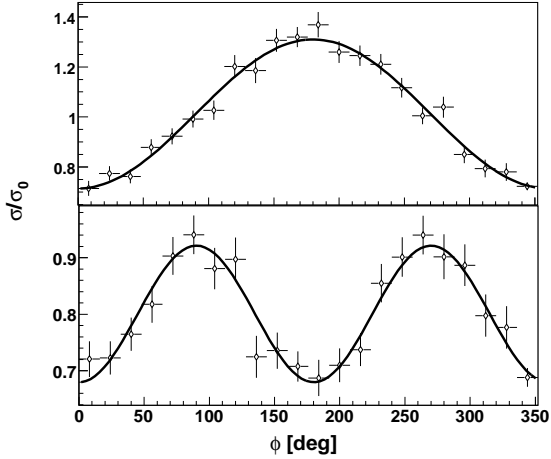}
\caption{The ratio of the spin-dependent cross
section to the unpolarized one for a pure vector-polarized deuteron
beam (top panel) and a pure tensor-polarized deuteron beam (bottom
panel) for ($\theta _{p}=28^\circ , \theta _{d}=30^\circ , \phi _{12}=180^\circ, S=210$~MeV).}
\label{ddBAsy}       
\end{figure}
\begin{sloppypar}
According to Eq.~\ref{ddbreakcrosFurmola}, for a deuteron beam with a
pure vector polarization, the ratio $\frac {\sigma}{\sigma _{0}}$ should show a
$\cos \phi$ distribution. When a pure tensor-polarized deuteron beam
is used, the ratio $\frac {\sigma}{\sigma _{0}}$ should show a $\cos
2\phi$ distribution.  These asymmetries are exploited to extract the 
vector-analyzing power, $iT_{11}$ and the tensor-analyzing powers, 
$T_{20}$ and $T_{22}$, for every kinematical configuration,
$(\theta_{p},\theta_{d},\phi_{12},S)$. 
\end{sloppypar}
\begin{figure}[!th]
\centering
\includegraphics[width=1.0\columnwidth,angle=0]{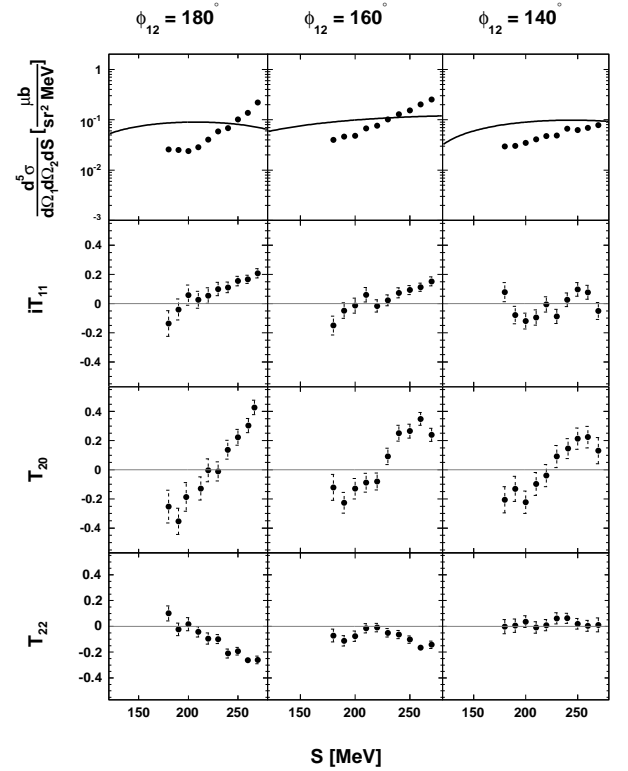}
\caption{The cross sections, vector-, and tensor-analyzing
 powers at $(\theta_{d}, \theta_{p}) = (15^{\circ}, 15^{\circ})$ as a
 function of $S$ for different azimuthal opening angles. The solid
 curves in the top panels correspond to phase-space distributions.
 They have arbitrary normalization with respect to the data.  The gray
 lines in other panels show the zero level of the analyzing powers.
 Only statistical uncertainties are indicated.}
\label{dd1}
 \end{figure}
\begin{sloppypar}
 Figure~\ref{ddBAsy} shows the
ratio $\frac {\sigma}{\sigma _{0}}$ for a pure vector-polarized
deuteron beam (top panel) and a pure tensor-polarized deuteron beam
(bottom panel) for ($\theta_{p}=28^\circ$, $\theta_{d}=30^\circ$,
$\phi_{12}=180^\circ$, $S=210$~MeV). The curves in the top and bottom
panels are the results of a fit based on Eq.~\ref{ddbreakcrosFurmola}
through the obtained asymmetry distribution for a beam with a pure
vector and  tensor polarization, respectively.  The amplitude of the
$\cos\phi $ modulation in the top panel equals $\sqrt{3}p_{Z}iT_{11}$
and the amplitude of the $\cos 2\phi $ modulation in the lower panel
equals $-
\frac {\sqrt{3}}{2}p_{ZZ}T_{22}$. The offset from 1 in the lower panel
equals $- \frac {1}{\sqrt{8}}p_{ZZ}T_{20}$.  The polarization values
have been measured independently using BINA and verified by
measurements using a Lamb-Shift polarimeter~\cite{RamazaniA_Ref31}, 
and were found to be $p_{Z} = -0.601 \pm 0.029$ and $p_{ZZ} = -1.517 \pm 0.032$.
\end{sloppypar}

\begin{sloppypar}
The differential cross section and vector- and tensor-analyzing powers
of a few kinematical configurations of the three-body break-up
reaction are obtained.  The differential cross sections
were compared with a phase-space distribution obtained from a
Monte Carlo simulation based on the GEANT3 framework. This comparison
demonstrates that there are large variations in the dynamic part of
the $t$-matrix as a function of $S$ for different configurations.
\end{sloppypar}

\begin{sloppypar}
Figure~\ref{dd1} represents the cross sections, vector-, and
tensor-analyzing powers at $(\theta_{d}, \theta_{p}) = (15^{\circ},
15^{\circ})$ as a function of $S$ for different azimuthal opening
angles. The solid curves in the top panels correspond to phase-space
distributions. They have arbitrary normalization with respect to the
data.  The gray lines in other panels show the zero level of the
analyzing powers.  Only statistical uncertainties are indicated.  The
total systematic uncertainty for the cross sections and analyzing
power are estimated to be $\sim 7\%$ and $\sim 4.5\%$, respectively.
\end{sloppypar}

\begin{sloppypar}
The three-body break-up reaction in deuteron-deuteron scattering has
been identified uniquely using the scattering angles, $\theta_{d},
\theta_{p}$, the energies and the TOF information of the scattered
particles. In this work, we analyzed a part of the data in which both protons
and deuterons  scattered into the forward wall of BINA.
The performed four-body scattering experiments with BINA and the BBS
provide an extensive database for the elastic and transfer channels at
65~MeV/nucleon and 90~MeV/nucleon and also for the three-body break-up
reaction at 65~MeV/nucleon. The available dataset for the
deuteron-deuteron scattering at intermediate energies can be used to
check the upcoming theoretical calculations for the four-body systems.
\end{sloppypar}

\end{document}